\documentclass[pts12]{revtex4}
\usepackage{graphics}
\usepackage{amssymb}
\usepackage{amsmath}

\newcommand{\rd}{\mathrm{d}}

\begin{document}

\author{Douglas Armstead$^{a}$}
\email{dna2@physics.umd.edu}
\affiliation{University of Maryland, College Park, Maryland 20904}
\author{Brian Hunt$^{b}$}
\affiliation{University of Maryland, College Park, Maryland 20904}
\author{Edward Ott$^{a,c}$}
\affiliation{University of Maryland, College Park, Maryland 20904}

\title{Anomalous Diffusion in Infinite Horizon Billiards}
\date \today

\begin{abstract}
We consider the long time dependence for the moments of displacement 
$\langle |r|^{q} \rangle$ of infinite horizon billiards, given a bounded 
initial distribution of particles.  For a variety of 
billiard models we find $\langle |r|^{q} \rangle \sim t^{\gamma_{q}}$
(up to factors of $\log t$).  The time exponent, $\gamma_{q}$, is piecewise 
linear and equal to $q/2$ for $q<2$ and $q-1$ for $q>2$.  We discuss the 
lack of dependence of this result on the initial distribution of particles 
and resolve apparent discrepancies between this time dependence and a prior 
result. The lack of dependence on initial distribution follows from 
a remarkable scaling result that we obtain for the time evolution of the 
distribution function of the angle of a particle's velocity vector.
\end{abstract}

\maketitle

\section{Introduction}

\begin{figure}
\begin{center}
\resizebox{120mm}{!}{\includegraphics{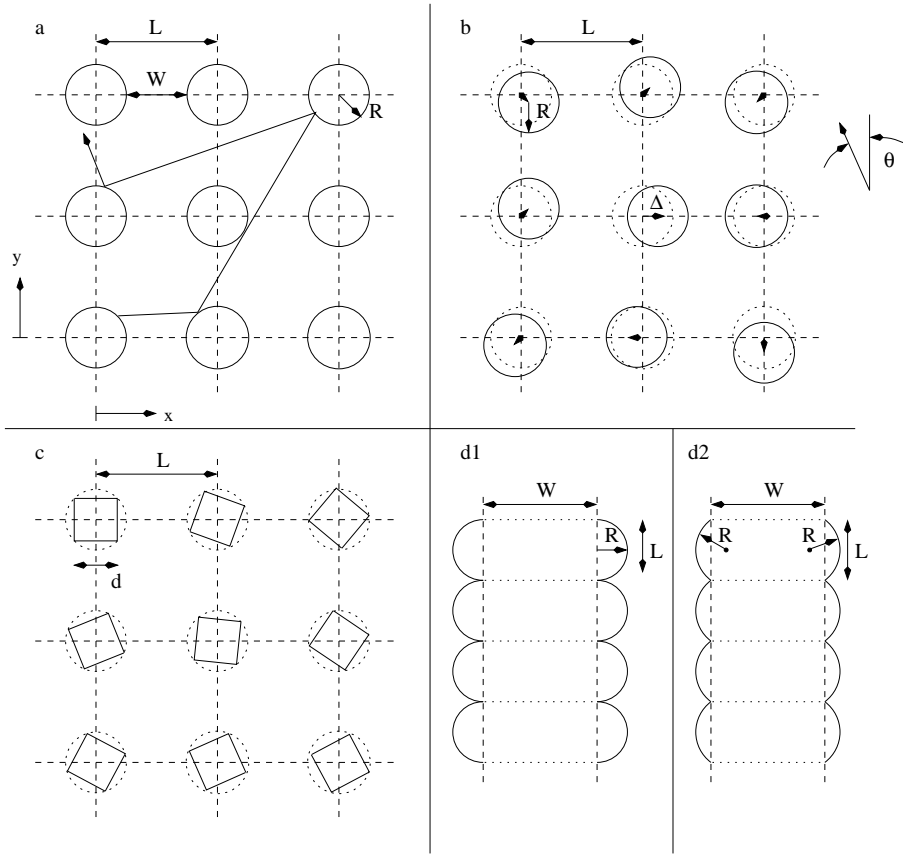}}
\caption{The four infinite horizon billiard structures that we consider 
include:(a) the Sinai billiard such that the channel width $W=L-2 R > 0$, 
(b) the Sinai billiard with random displacements $\Delta$ away from the 
square matrix such that $W=L-2( R+ \Delta) >0$, (c) randomly oriented squares 
such that $W=L-\sqrt{2}d>0$, (d) scalloped channel with (d1) semicircular arcs 
and (d2) arcs subtending an angle less than $180^{o}$.}
\label{fig:examples}
\end{center}
\end{figure}

Diffusion of particles in an infinite domain billiard is a well studied 
problem [1-5,9,14].  By a billiard we refer to the motion of a point particle 
in a two dimensional domain in which the particle moves with constant 
velocity in straight 
line orbits executing specular reflection (i.e., angle of incidence equals 
angle of reflection) from fixed boundaries.  By an infinite domain we 
refer to an unbounded two dimensional region.  An early 
consideration of diffusion in a billiard as a model in physics was made by 
Lorentz [1] to model electrons in a metal.  In this model (called the Lorentz 
gas) particles move freely and reflect specularly from fixed, randomly
placed, hard-wall scatterers. A modification of the two dimensional Lorentz 
gas in which there are circular scatterers on a square 
lattice is an example of an infinite horizon billiard, called the Sinai 
billiard [5], and is illustrated in Fig.~\ref{fig:examples}(a).  
Infinite 
horizon billiards are the subset of infinite domain 
billiards that contain channels through which a particle 
may pass without ever reflecting off a billiard wall.  

\begin{figure}
\includegraphics{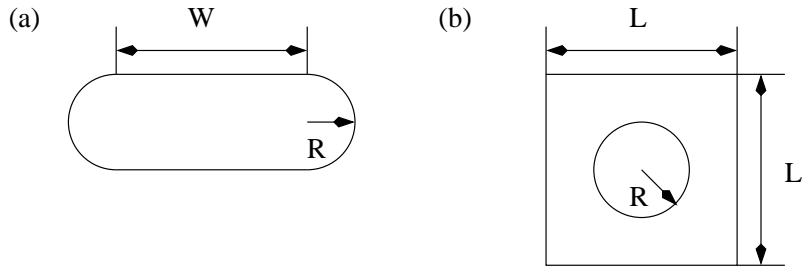}
\caption{(a)The stadium billiard has a domain bounded by two 
semi-circular arcs of radius $R$ and two straight lines of     
length $W$.  Particle motion in this billiard can be mapped
to motion in the scalloped channel [Fig.~\ref{fig:examples}(d1)].  
In a similar manner particle motion in the bounded domain billiard
shown in (b) can be mapped to motion in the infinite domain billiard of
Fig.~\ref{fig:examples}(a).}
\label{fig:stadium}
\end{figure}

In this paper we consider diffusion in infinite horizon billiards.  The examples that we 
study numerically are shown in Figs.~\ref{fig:examples}(a)-(d).  
The billiards in Fig.~\ref{fig:examples} include: (a) the Sinai billiard, 
composed of circular, hard wall scatterers arranged on a square lattice 
such that the scatterers do not touch each other; (b) a modification of model 
(a) in which the circular scatterers are 
randomly displaced (random in direction and magnitude) by at most 
$\Delta < L/2-R$, so that there are channels of width $L-2(R+\Delta)$ 
accommodating free motion; (c) randomly oriented 
square scatterers on a square lattice; and (d) the scalloped channel, in which
the domain is infinite in the $y$ direction and bounded in the $x$ direction
by circular arc segments, each subtending an angle less than or equal to 
$180^{o}$.  Figure~\ref{fig:examples}(d1) shows the case of the scalloped 
channel where the circular arc segments are semi-circles, while 
Fig.~\ref{fig:examples}(d2) shows the case where the arcs subtend an angle 
less than $180^{o}$.
Particle motion for the situation in Fig.~\ref{fig:examples}(d1) is 
equivalent to particle motion for a stadium-type billiard 
[see Fig.~\ref{fig:stadium}(a)]; the particle 
motion within the scalloped channel can be folded into the stadium billiard 
via reflection of the particle at a straight wall as it passes to the next 
cell.  
By cells we mean each portion of the scalloped channel domain between 
the dotted lines of Fig.~\ref{fig:examples}(d).  In a similar manner 
particle motion in the bounded billiard of Fig.~\ref{fig:stadium}(b) can 
be thought of as equivalent to motion in the infinite billiard of 
Fig.~\ref{fig:examples}(a).

One important means of characterizing transport in an infinite domain 
two dimensional billiard is through the phase space probability 
distribution function (pdf), $P(x,y,\theta,t)$, where $\theta$ is the angle 
of the particle velocity and we take all 
particle velocities to have magnitude one.  From the pdf one can calculate the 
displacement moments of the distribution.   The $q$th moment at time $t$ is 

\begin{equation}
\langle r^{q} \rangle =
\langle (x^{2}+y^{2})^{q/2} \rangle
	=\int_{0}^{2\pi}\int_{\mathcal{R}}
	(x^{2}+y^{2})^{q/2}P(x,y,\theta,t) \rd{x}\rd{y} 
	\frac{\rd \theta}{2\pi},
\end{equation}
where $r=\sqrt{x^{2}+y^{2}}$ and $\mathcal{R}$ is the (infinite) spatial 
domain of the billiard and where throughout this paper we take the initial
pdf, $P(x,y,\theta,0)$, to be bounded, $|P(x,y,\theta,0)|<K$, and to be zero 
outside some finite region.
For the infinite horizon billiards shown in Fig.~\ref{fig:examples} 
we find that the moments of the displacement have a time dependence
\begin{equation} \label{eq:moment_intro}
\langle r^{q} \rangle \sim t^{\gamma_{q}},
\end{equation}
which we use as shorthand for 

\begin{equation}\label{eq:shorthand}
\gamma_{q}=\lim_{t \to \infty} \frac{\log \langle r^{q} \rangle}
				{\log t}.
\end{equation}
For all cases in Fig.~\ref{fig:examples} we find the exponent $\gamma_{q}$ to 
be 

\begin{equation} \label{eq:gamma}
\gamma_{q}= \left\{
\begin{array}{ll}
q/2 & q<2 \\
q-1 & q>2
\end{array} \right. .
\end{equation}
Results of the form (\ref{eq:moment_intro}) with $\gamma_{q}$ composed of 
piecewise linear functions (different from (\ref{eq:shorthand})) have
also been obtained in other situations of Hamiltonian transport [6, 7].
The occurrence of an exponent $\gamma_{q} \neq q/2$ is commonly referred to as
\textit{anomalous diffusion}, and $\gamma_{q} > q/2$ 
$(\gamma_{q} < q/2)$ is called superdiffusion (subdiffusion).

\begin{figure}
\resizebox{120mm}{!}{\rotatebox{-90}{\includegraphics{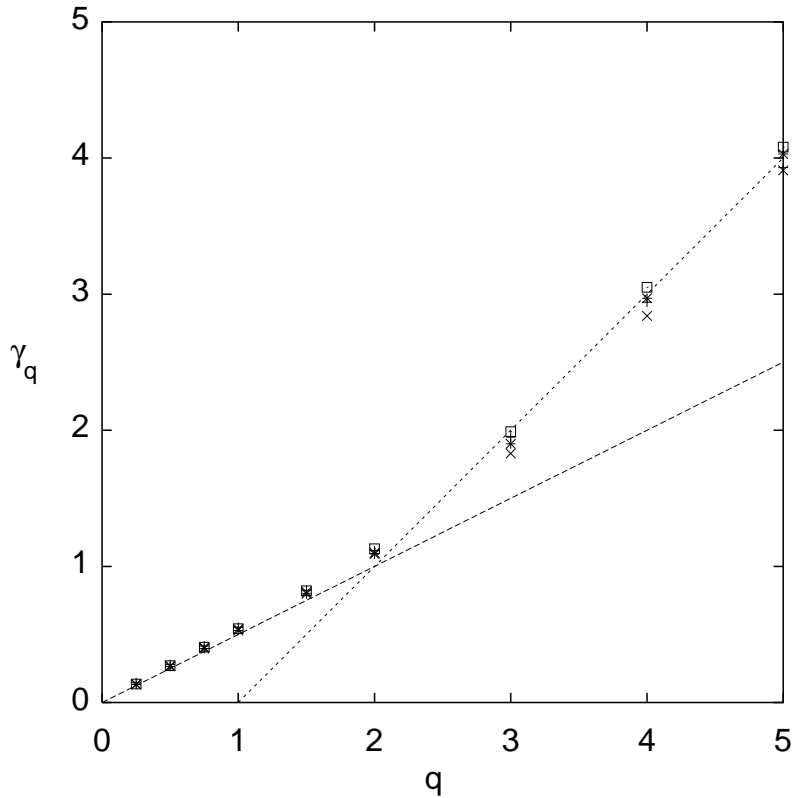}}}
\caption{The dependence of the $q$th moment, $\langle |r|^{q} \rangle$, on 
time for the 
models discussed in this paper.  Shown is the numerically determined time 
exponent $\gamma_{q}$; with the symbols $\times$, $\Box$, $*$, and + 
corresponding to the billiards depicted in 
Figs.~\ref{fig:examples}(a), \ref{fig:examples}(b), 
\ref{fig:examples}(c), and \ref{fig:examples}(d1) respectively.  
The dashed line represents $\gamma_{q}=q/2$ and 
the dotted line represents $\gamma_{q}=q-1$.}
\label{fig:diffusion}
\end{figure}
Figure~\ref{fig:diffusion} shows the results of numerical experiments testing
(\ref{eq:gamma}).  In these numerical experiments we start with a cloud of 
many initial conditions distributed uniformly in the accessible space 
occupied by one cell [the region outside of the scatterer and within 
$0 \leq y \leq L, 0\leq x \leq L$ for Figs.~\ref{fig:examples}(a)-(c),
and $0 \leq y \leq L$ for Figs.~\ref{fig:examples}(d)] and uniformly in 
$0 \leq \theta \leq 2\pi$.  We then evolve the orbit of each particle in the 
cloud forward in time [8] and obtain $\langle r^{q} \rangle$.  In all cases 
the results we obtain are consistent with (\ref{eq:gamma}).  The motivation 
for choosing an initial distribution uniform in space and angle is that 
motion of an 
orbit from a typical initial condition in the billiards of 
Fig.~\ref{fig:stadium} is known to be ergodic, generating an invariant 
density uniform in space and in angle $0 \leq \theta \leq 2\pi$.
In another set of numerical experiments we used initial conditions distributed 
uniformly in intervals of $\theta$ such that the lengths of the initial 
flights are bounded [e.g., for Figs.~\ref{fig:examples}(a)-(c), 
$\theta_{0} \leq \theta \leq \pi/2-\theta_{0}$, 
$\pi/2+\theta_{0} \leq \theta \leq \pi-\theta_{0}$, 
$\pi+\theta_{0} \leq \theta \leq 3\pi/2-\theta_{0}$,
$3\pi/2+\theta_{0} \leq \theta \leq 2\pi-\theta_{0}$ where 
$\theta_{0} < \pi/4$].
Again, agreement with (\ref{eq:shorthand}) and (\ref{eq:gamma}) was found. 
This indicates that the angular particle distribution relaxes to the 
uniform distribution sufficiently fast that the results (\ref{eq:shorthand}) 
and (\ref{eq:gamma}) are not modified.  Explaining the reason for this 
insensitivity to the initial distribution is one 
of the main contributions of this paper (Sec.~2C).  Our other main 
contribution is the result
that, for long time, initial distributions with no particles in a $\theta$ 
interval about a direction of infinitely long flight (say $\theta=0$) lead
to long-time distributions with an invariant scaling form.  Specifically, 
$\hat{P}(\theta,t)=\int_{\mathcal{R}}P(x,y,\theta,t)\rd x \rd y$ approaches 
$\tilde{P}(\phi)$, where $\phi=\theta t$ and $\tilde{P}(\phi)$ is independent 
of $t$.  Furthermore, the same 
$\tilde{P}(\phi)$ universally applies for the billiards of 
Figs.~\ref{fig:examples}(a), \ref{fig:examples}(b), \ref{fig:examples}(c), 
\ref{fig:examples}(d2), but $\tilde{P}(\phi)$ is different for the billiard of 
Fig.~\ref{fig:examples}(d1).  We use this scaling result to show the 
insensitivity of (\ref{eq:gamma}) to the initial particle distribution.
 
Bleher [9] showed for infinite horizon billiards that the 
\textit{limit in distribution} as $t \to \infty$ of the pdf of the particle 
displacements in the billiard is Gaussian with a width that increases with 
time as $\sqrt{t \log t}$.  (For ordinary diffusion, the result is the same 
except that the width increases as $\sqrt{t}$.)  However, the asymptotic $t$ 
dependence of the moments cannot be calculated from Bleher's result.  
This is discussed in Sec.~3.
Note that the definition of the symbol $\sim$ given in 
Eqs.~(\ref{eq:moment_intro})~and~(\ref{eq:shorthand}) is such that logarithmic 
corrections to the scaling of $\langle r^{q}\rangle$ with 
time are not included [e.g., if $\langle r^{2} \rangle \cong C t \log t$, 
where $C$ is a constant, as suggested 
by [9], then $\gamma_{2}$ from (\ref{eq:shorthand}) is one, consistent with 
(\ref{eq:gamma})].

\section{Theory}

The difference between particle transport in an infinite horizon billiard 
and particle transport in the case of normal diffusion is due to the 
arbitrarily long trajectories found in the infinite horizon billiard.
These long flights occur in the channels between the 
scattering boundaries of the billiard.  When a particle is traveling nearly
parallel to the axis of the channel [for 
Figs.~\ref{fig:examples}(a)-(c) there are many channels parallel to both the 
$x$ and 
$y$ axes whereas for Fig.~\ref{fig:examples}(d) there is only one 
channel, which is parallel to the $y$ axis], it will travel long 
distances between reflections off the billiard wall.  Let $\theta_{n}$ be 
the angle between the trajectory of a particle and the axis of 
the channel after the $n$th reflection of that particle with a billiard wall.
The length $r_{n}$ of the flight is of the order of $1/|\theta_{n}|$ for 
$|\theta_{n}| \ll 1$.  

For some of the billiards we consider there are strong correlations 
between the $\theta$ values from
one reflection to the next.  This is especially true of the scalloped 
channel with semicircular arcs (stadium billiard).  Here, upon reflection,
the angle $\theta_{n}$
can change by, at most, a factor of three [12]; 
$|\theta_{n}|/3 \leq |\theta_{n+1}| \leq 3 |\theta_{n}|$.  Thus if 
$|\theta_{n}|$ 
is small, $|\theta_{n+1}|$ is also small, and both represent long flights.  
This suggests an extreme model for the particle transport in which the length 
of a 
flight for each particle does not change from reflection to reflection; the 
length of each flight is completely correlated with the previous flight, but
different particles have different flight lengths.

\subsection{The Completely Correlated Model}

Consider a one dimensional system in which an ensemble of particles executes
a random walk.  The particles in the ensemble differ from each other in the 
length of the step, $\Delta r_{\xi}$, each particle takes.  
\begin{equation} \label{eq:xi_dist}
\Delta r_{\xi}=1/ \xi, \ \xi \in [0,1].
\end{equation}
The random variable $\xi$ is distributed uniformly in $[0,1]$ ($\xi$ is 
inspired by $\theta_{n}$ defined above, but is chosen to occupy the
interval $[0,1]$ for simplicity), and $\xi$ for a particle does not 
change from step to step (complete correlation).  Since every billiard 
particle has the same speed we normalize the magnitude of the velocity 
to 1 and thus take the time between steps in our one dimensional random 
walk model to be
\begin{equation}
\Delta t_{\xi}=1/ \xi.
\end{equation}
Since the random walk behaves like normal diffusion when the number of 
steps is large, the 
$q$th moment for each $\xi$, 
$\left\langle |r|^{q} \right\rangle _{\xi}$, will be
\begin{equation}
\langle |\frac{r}{\Delta r_{\xi}}|^{q} \rangle _{\xi} \sim 
(\frac{t}{\Delta t_{\xi}})^{q/2} \textrm{, for } t \gg \Delta t_{\xi}.
\end{equation}
Substitution for $\Delta r_{\xi}$ and $\Delta t_{\xi}$ yields, 
for long time $t$, $(t \gg  \Delta t_{\xi})$
\begin{equation} \label{eq:moment_species}
\langle |r|^{q} \rangle _{\xi}\sim(t/ \xi)^{q/2} 
\end{equation}
for each $\xi$.  

To find the $q$th moment for the entire ensemble one 
needs to take an average over all particles:
\begin{equation}
\langle |r|^{q} \rangle =\int_{0}^{1} \langle |r|^{q} \rangle _{\xi} 
 \rd \xi.
\end{equation}
This average has two main contributions:
\begin{equation}\label{eq:contributions}
\langle |r|^{q} \rangle  \sim \int_{N/t}^{1} (t/\xi)^{q/2} 
\rd \xi+\int_{0}^{1/t}t^{q} \rd \xi
\end{equation}
where $N$ is chosen to be large, so that the random walks have made many steps.
The first term in Eq.~(\ref{eq:contributions}) represents the fraction of 
particles that have executed at least $N$ steps and so can be described by 
Eq.~(\ref{eq:moment_species}).  The second term comes from those particles 
that are still in their first flight (with velocity = 1).  
The contribution from the interval $N/t \geq \xi \geq 1/t$ 
[omitted in (\ref{eq:contributions})] has neither the majority of walkers 
(in the limit of long time) nor the most 
extreme displacements and so does not dominate the other terms.  For long time 
Eq.~(\ref{eq:contributions}) yields

\begin{equation}\label{eq:result}
\langle |r|^{q} \rangle \sim \left\{
\begin{array}{ll}
t^{q/2} & q \leq 2 \\
t^{q-1} & q \geq 2
\end{array} \right. .
\end{equation}

This simple toy model shows that it is the very long flights allowed by the 
open channels that give rise to the non-Gaussian behavior for moments greater
than 2.  One might then think that for the real billiard systems 
of Fig. \ref{fig:examples}  the behavior for $q>2$ is a trivial result of 
the initial distribution of angles.  However, this is not the case and the 
behavior (\ref{eq:result}) is robust to changes in the initial distribution 
of particles.  As discussed in Sec.~2C, non-uniform initial
angular distributions scatter rapidly enough that (\ref{eq:result}) still 
holds even if the initial distribution has no particles traveling 
nearly parallel to the channels.

\subsection{Moment Equation for Infinite Horizon Billiards with a Uniform 
Initial Angular Distribution}

We now consider a uniform initial spatial (within a cell) and 
angular distribution and for the case of the ergodic billiards of 
Figs.~\ref{fig:examples}(a) (Sinai billiard) and \ref{fig:examples}(d) 
(scalloped billiard).  In these cases such a distribution is stationary when 
$(x,y)$ is taken modulo the appropriate cell period.  We show that 
$\gamma_{q}$ is given by Eq.~(\ref{eq:gamma})  for these real billiard 
systems.  The particle transport 
in an infinite horizon billiard must proceed at least as quickly as a 
random walk process (i.e., as fast as normal diffusion).  While there exist
mechanisms for faster than diffusive transport (to be discussed below), 
there is no stable mechanism to stop or trap a billiard particle.  The 
periodic orbits that might trap the particle in these systems are all 
exponentially unstable.  That a random walk is a lower bound on the 
transport in the billiard system implies that $\gamma_{q} \geq q/2$.  

In addition to diffusive like behavior, particle trajectories that consist 
of a single long flight (``ballistic'' flight) also participate in 
particle transport.  If every particle were to move ballistically, we would 
find $\gamma_{q}=q$.  This provides an upper bound on $\gamma_{q}$; not every 
particle will execute a single uninterrupted flight.  
We can, however, place a lower bound on the fraction of particle that do.
Since the distribution of particle velocities is uniform in orientation, the 
fraction of particles executing flights of distance $vt$ or more 
(where $v$ is the velocity of the particle) involve a 
fraction of order of $W/t$ of the particles occupying a 
channel at any one time, where $W$ is the channel width defined in 
Fig.~\ref{fig:examples}.  (Although we have defined $v=1$, we 
retain $v$ in this section to clarify when we are speaking of distances and 
when we are speaking of times.)  The contribution to 
$\langle |r|^{q} \rangle$ from these particles is

\begin{equation}\label{eq:lower_bound}
\langle |r|^{q} \rangle \gtrsim (vt)^{q}(C/t)=Ct^{q-1},
\end{equation}
where $C$ is a constant depending on the size of the channel.

There are two points that can be fixed on the graph of $\gamma_{q}$ vs. $q$.  
The first is the zeroth moment, which by the conservation of probability 
is identically equal to one.  Thus $\gamma_{0}=0$.  The second 
point is fixed by the diffusion coefficient, which relates the second 
moment to time.  It has been found [14] that 
$\langle r^{2} \rangle$ is of the order $t \log t$ and so, consistent 
with the definition (\ref{eq:shorthand}), 
ignoring factors of $\log t$ we have $\gamma_{2} = 1$.  

Next we argue that $\gamma_{q}$ is a concave up function of $q$.
Invoking the Cauchy-Schwartz inequality 
$(||(x \cdot y)|| \leq ||x|| \times ||y||)$, for $ 0 \leq \epsilon \leq q$ 

\begin{equation}
\langle |r|^{q} \rangle =
\langle |r|^{(q+\epsilon)/2}|r|^{(q-\epsilon)/2} \rangle \leq 
\langle |r|^{q+\epsilon} \rangle ^{1/2} 
\langle |r|^{q-\epsilon} \rangle ^{1/2}.
\end{equation}
Thus

\begin{equation} \label{eq:concavity_condition}
\log \langle |r|^{q} \rangle \leq \frac{\log \langle |r|^{q+\epsilon} 
\rangle + \log \langle |r|^{q-\epsilon} \rangle} {2}.
\end{equation}
From (\ref{eq:moment_intro}), (\ref{eq:shorthand}), and 
(\ref{eq:concavity_condition}) we have that the graph of $\gamma_{q}$ 
versus $q$ is indeed concave up
\begin{equation} \label{eq:gamma_condition}
\gamma_{q} \leq \frac{\gamma_{q+\epsilon}+\gamma_{q-\epsilon}}{2}.
\end{equation}

Finally, we show that Eq.~(\ref{eq:gamma}) holds, i.e., that $\gamma_{q}=q/2$
for $q \leq 2$ and $\gamma_{q}=q-1$ for $q \geq 2$.
We have given lower bounds on $\gamma_{q}$, i.e. $\gamma_{q} \geq q/2$ and
$\gamma_{q} \geq q-1$.  
We have also pinned the value of $\gamma_{q}$ at two values: $\gamma_{0}=0$
and $\gamma_{2}=1$.  
The first lower bound, the two known values of $\gamma_{q}$, and the 
concavity condition (\ref{eq:gamma_condition}) force $\gamma_{q}=q/2$ 
for $q \leq 2$.  
There is a trivial upper bound on $\gamma_{q}$, $\gamma_{q} \leq q$.  This 
upper bound, along with Eq.~(\ref{eq:concavity_condition}) implies that the 
slope of $\gamma_{q}$ 
never exceeds one.  Since $\gamma_{2}=1$, $\gamma_{q}\leq q-1$ for $q \geq 2$.
Therefore, for $q \geq 2$ both the upper bound and the lower bound coincide 
resulting in $\gamma_{q}=q-1$ for $q \geq 2$.

\subsection{Non-uniform Initial Angular Distributions}

The discussions of the previous two subsections relied on the existence of 
a uniform distribution for $\theta_{n}$ in the particle ensemble.  This 
applies if one starts with a distribution uniform in angle and uniform in 
the space within a cell, in which case the particle ensemble will retain the 
uniform angular distribution.  On the other hand, even if we have an 
initial distribution that is non-uniform, it will (under very general 
conditions) relax to the uniform distribution as time increases.  The 
question we now address is whether this relaxation is fast enough to yield 
the same result for the temporal scaling of $\langle | r^{q}|\rangle$,
Eq.~(\ref{eq:result}), as the initially uniform distribution.

To consider this question, we first study the relaxation of an initial 
particle distribution with no particles in a finite gap around the channel 
direction.  For example, for the scalloped channel we consider the case 
where the initial particle distribution is uniform in the space within 
a cell $\mathcal{C}$ and

\begin{equation} \label{eq:scallopedInit}
P(x, y, \theta, t=0)= \left\{
\begin{array}{ll}
0 & \textrm{for } |\theta|<\theta_{o} \textrm{ or } |\theta-\pi|<\theta_{o}, \\
K & \textrm{ otherwise}
\end{array} \right. ,
\end{equation}
where $0 \leq \theta \leq 2\pi$ is the angle of the particle velocity vector 
with the $y$-axis, $K$ and $\theta_{o}$ are constants, and $(x, y)$ is in 
$\mathcal{C}$.  Similarly, for 
the cases of Figs.~\ref{fig:examples}(a)-\ref{fig:examples}(c) we consider 
an initial distribution, 

\begin{equation}\label{eq:initCond}
P(x, y, \theta, t=0)= \left\{
\begin{array}{ll}
0 & \textrm{for } |\theta|, |\theta-\pi/2|, |\theta-\pi|, |\theta-3\pi/2|<\theta_{o},\\
K & \textrm{otherwise}
\end{array} \right. .
\end{equation}
In particular, we focus on the behavior of $P(x, y, \theta, t)$ for $|\theta|$ 
small, $t$ large and $(x,y)$ in a channel.  (Similar results apply for 
(\ref{eq:initCond}) with $|\theta-\pi/2|$ or 
$|\theta-\pi|$ or$|\theta-3\pi/2|$ small.)  For all cases shown in 
Fig.~\ref{fig:examples}, we find a remarkable scaling behavior.  Let 
$\hat{P}(\theta,t)=\iint P \rd x \rd y$ where the spatial integral is over a 
channel.  Then, if 
we introduce the scaled variable $\phi=\theta t$, we find that, in all 
the cases we have tested, the angular distribution function 
$\hat{P}(\theta,t)$ approaches a stationary form.  That is,

\begin{equation}
\hat{P}(\theta, t) \to \tilde{P}(\phi) \textrm{ as } t \to \infty.
\end{equation}

\begin{figure}
\resizebox{120mm}{!}{\includegraphics{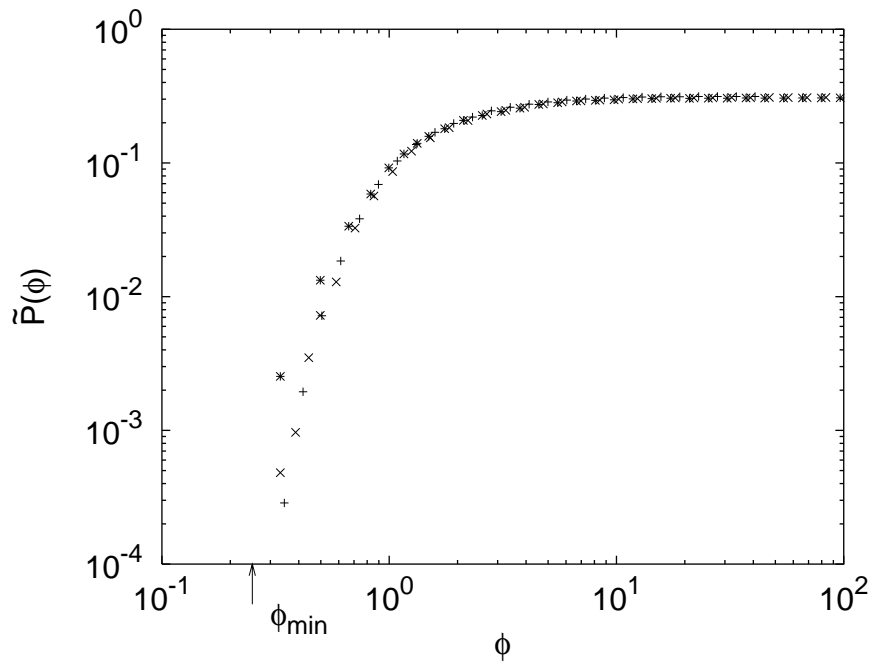}}
\caption{The time evolution of an initially uniform angular distribution 
of particles in the interval $[.3,\pi /2]$ for the scalloped channel with 
semi-circular arcs.  The distribution is shown at 
three times; + at $t=27$, $\times$ at $81$, and $*$ at $243$.  The 
distribution of particles in the channel of width $W$ becomes 
self-similar and static in the coordinate 
$\phi=\theta t$ with $\tilde{P}(\phi)=0$ for $\phi < \phi_{min}$.}
\label{fig:spread}
\end{figure}
We illustrate this with a numerical calculation on the scalloped billiard 
with $180^{o}$ arcs (Fig.~\ref{fig:examples}(d1)) in Fig.~\ref{fig:spread}.  
In generating this figure we follow the evolution of a large number of orbits 
initialized according to Eq.~(\ref{eq:scallopedInit}), and form the 
distribution $\hat{P}(\theta,t)$ using a histogram approximation.  As shown,
at successively larger $t$ the distribution approaches a time independent 
form $\tilde{P}(\phi)$ where $\tilde{P}(\phi)$ is zero for 
$\phi < \phi_{\textrm{min}}$, and, as $\phi$ increases past 
$\phi_{\textrm{min}}$, $\tilde{P}(\phi)$ increases, asymptoting to a constant 
for large $\phi$.

\begin{figure}
\resizebox{120mm}{!}{\includegraphics{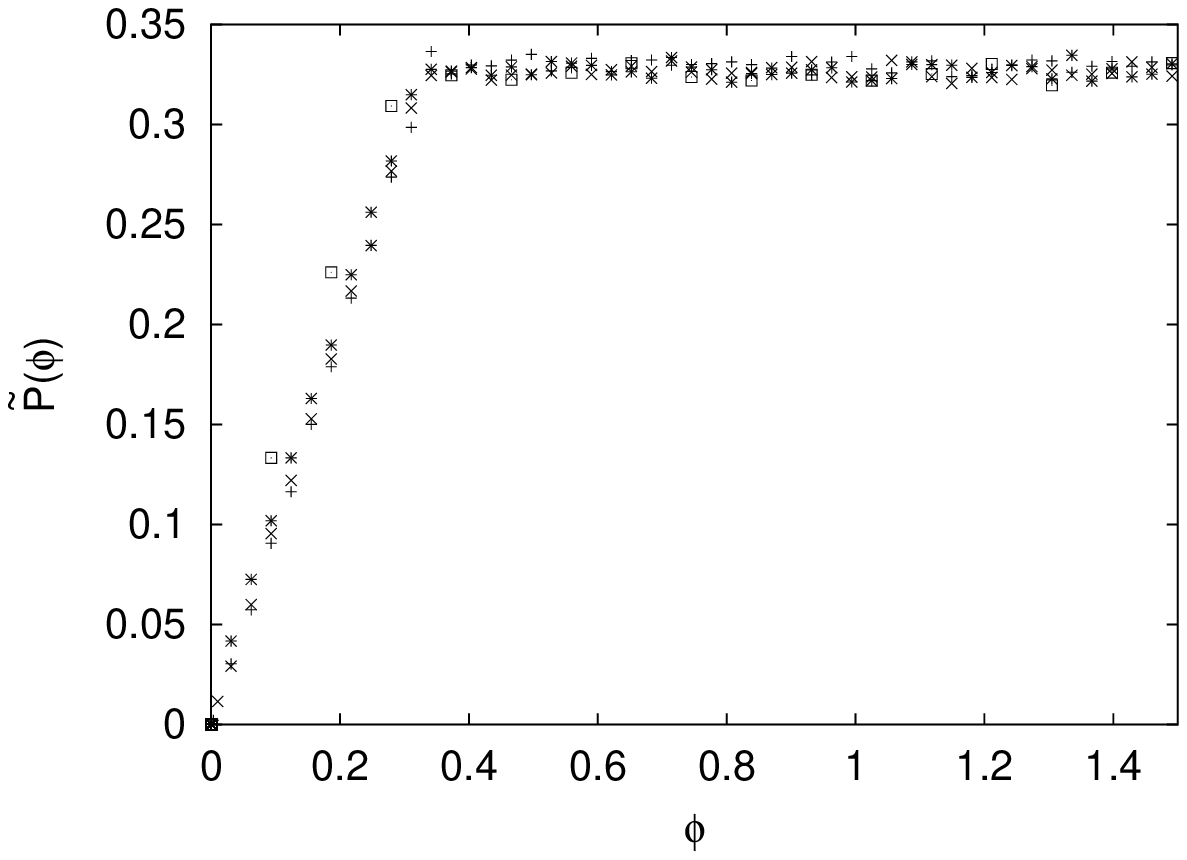}}
\caption{The time evolution of an initially uniform angular distribution 
of particles in the $\theta$ interval $[.3,\pi /2]$ for the Sinai billiard 
with $L/R=1$. The distribution is shown at 
four times; + at $t=9$, $\times$ at $27$, $*$ at $81$, and $\square$ at 
$243$.  The distribution of particles in a channel becomes self-similar 
and static in the coordinate $\phi=\theta t$.}
\label{fig:sinai}
\end{figure}
A similar numerical experiment for the Sinai billiard 
(Fig.~\ref{fig:examples}(a)) yields the result in Fig.~\ref{fig:sinai}.  In
this case the large time distribution assumes the form,

\begin{equation} \label{eq:Ptilde}
\tilde{P}(\phi)= \left\{
\begin{array}{ll}
C(\phi/\phi_{o}) & \textrm{ for } 0<\phi<\phi_{o},\\
C & \textrm{ for } \phi > \phi_{o},
\end{array} \right. ,
\end{equation}
where $C$ and $\phi_{o}$ are constants.  Furthermore, we numerically obtain
this same form for all the other cases of Fig.~\ref{fig:examples}, [except 
for the case of the scalloped billiard with $180^{o}$ arcs 
(Fig.~\ref{fig:examples}(d1)) which gives the result in Fig.~\ref{fig:spread}].
We explain the reason for the result (\ref{eq:Ptilde}) and why it does not
apply for the billiard of Fig.~\ref{fig:examples}(d1) subsequently, but before
doing that we first show that these results for $\tilde{P}(\phi)$ imply the 
applicability of Eq.~(\ref{eq:gamma}) for the large time behavior of 
$\langle |r|^{q}\rangle$.

In order to show that (\ref{eq:gamma}) applies consider the fraction 
$\Xi (t_{o})$ of particles that, at some large time $t_{o}$, have 
$\phi<\phi_{*}$, where we take $\phi_{*}=2\phi_{\textrm{min}}$ for the case of 
Fig.~\ref{fig:examples}(d1) and $\phi_{*}=\phi_{o}$ for the other cases.  
Noting that $\phi<\phi_{*}$ implies $\theta<\phi_{*}/t$, we have that for 
large $t_{o}$

\begin{equation}
\Xi(t_{o}) \cong \int_{0}^{\phi_{*}/t_{o}} 
\tilde{P}(\theta t) \rd \theta \\
=(t_{o})^{-1} \int_{0}^{\phi_{*}} \tilde{P}(\phi) \rd \phi 
= K t_{o}^{-1},
\end{equation}
where $K$ is a constant.
Between times $t_{o}$ and $2t_{o}$ these particles experience flights of 
length $\sim vt_{o}$.  Hence for $t=2t_{o}$ these flights give a 
contribution to 
$\langle r^{q} \rangle$ that is approximately $(2K/t)(vt)^{q} \sim t^{q-1}$.
Thus the lower bound (\ref{eq:lower_bound}) still applies, and, by the 
reasoning in Sec.~2B, we again obtain (\ref{eq:gamma}).

We note that the asymptotic time dependence $\tilde{P}(\phi)$ with 
$\phi \sim \theta t$
is marginal in the sense that, if the repopulation of an initially empty 
channel were slower (in the sense below), then (\ref{eq:gamma}) would not be
recovered.  In order to see this, consider the hypothetical case where an 
asymptotic distribution $P(\phi)$ of the form in Fig.~\ref{fig:spread} or 
Fig.~\ref{fig:sinai} is 
still approached, but with a self-similar $\phi$ scaling given by 
$\phi_{\alpha} \sim \theta t^{\alpha}$.  We have already considered the 
case $\alpha=1$, while $\alpha<1$ ($\alpha > 1$) corresponds to slower 
(faster) filling in of the channel.  If, for $0< \alpha <1$ (i.e. slow 
repopulation) we pursue the same reasoning as above for the $\alpha=1$ case,
then we obtain the bound $\gamma_{q} \geq \alpha(q-1)$, and we can no longer
conclude that (\ref{eq:gamma}) holds.  For $\alpha >1$, we consider at time
$t_{0}$ a range of angles 
$\theta_{+} 
\sim 1/t_{o} >\theta > \theta_{-} 
\sim \phi_{*}/t_{o}^{\alpha}$.
Since $\alpha > 1$, we have that $\theta_{+} \gg \theta_{-}$, and replacing
$\theta_{-}$ by zero does not alter the estimate for the contribution to 
$\langle r^{q} \rangle$ from $\theta_{+} > \theta > \theta_{-}$.  Thus the 
lower bound estimate of Sec.~2B still applies for $0 < \alpha < 1$.
 
We now discuss the asymptotic forms $\tilde{P}(\phi)$ illustrated in 
Fig.~\ref{fig:spread} for the scalloped billiard with $180^{o}$ arcs, and 
in Fig.~\ref{fig:sinai} and Eq.~(\ref{eq:Ptilde}) for the other cases.

First we discuss the scalloped billiard with $180^{o}$ arcs.  A full 
treatment of the theory yielding $\tilde{P}(\phi)$ in this case will 
be given elsewhere [13]; in the present paper, we will limit our discussion 
to the basic reason for the difference between this case and the other cases.
In particular, we discuss why, for this case, $\tilde{P}(\phi)=0$ in a 
finite interval about zero, $\phi<\phi_{\textrm{min}}$.  From 
Fig.~\ref{fig:examples}(d1) we see that after a long flight in the channel
a particle will collide with the channel wall close to one of the cusp points
where two arcs touch.  For the case of 
$180^{o}$ arcs, the tangent to such a section of the channel wall is nearly 
horizontal.  Thus, upon reflection 
$\hat{\theta} \equiv \textrm{min}(|\theta|, |\theta-\pi|)$ will still be small.
In fact as shown in [12] by consideration of the geometry, 
$\hat{\theta}_{n+1}$ on the $(n+1)$st reflection cannot change by more than a 
factor of 3 from $\hat{\theta}_{n}$,

\begin{equation}
\hat{\theta}_{n}/3 \leq \hat{\theta}_{n+1} \leq 3\hat{\theta}_{n}, 
\textrm{ for } \hat{\theta}_{n} \ll 1.
\end{equation}
Thus, if $\hat{\theta}_{n}$ is small, $\hat{\theta}_{n+1}$ is still relatively
small.  We can obtain the lower bound on $\phi$ by considering the most extreme
case where $\hat{\theta}$ always decreases by 3 on every bounce,

\begin{equation}\label{eq:THETAhat}
\hat{\theta}_{n+1}=\frac{1}{3}
\hat{\theta}_{n},
\end{equation}

\begin{equation}\label{eq:time}
t_{n+1}=t_{n}+W/(v\hat{\theta}_{n}),
\end{equation}
where $W$ is the channel width defined in Fig.~\ref{fig:examples}(d1), $v$
is the particle velocity, and $t_{n}$ is the time of the $n$th reflection.
Multiplying (\ref{eq:THETAhat}) by (\ref{eq:time}) we have

\begin{equation}
\hat{\phi}_{n+1}=\frac{1}{3}\hat{\phi}_{n}+W/v,
\end{equation}
which, for large $n$, asymptotes to the solution $\hat{\phi}=3W/(2v)$.
Thus $\phi_{\textrm{min}}=3W/(2v)$, which agrees with our numerical solution
Fig.~\ref{fig:spread} (see also [13]).  

\begin{figure}
\begin{center}
\resizebox{120mm}{!}{\includegraphics{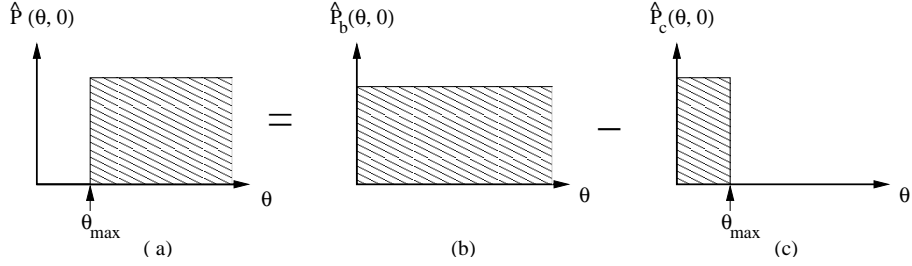}}
\caption{An initial distribution of particles with trajectories 
uniform in the angle $\theta$ for $\theta > \theta_{\textrm{max}}$ and zero 
for angles $\theta < \theta_{\textrm{max}}$, Fig.~(a), is equivalent to
a distribution uniform for all $\theta$, Fig.~(b), minus a distribution 
that is uniform for  $\theta < \theta_{\textrm{max}}$ and zero for 
$\theta > \theta_{\textrm{max}}$, Fig.~(c).  We use the time evolution of the 
distributions in Figs.~(b) and (c) to find the time evolution of Fig.~(a), see 
Fig.~\ref{fig:complement}.}
\label{fig:initDist}
\end{center}
\end{figure}

\begin{figure}
\includegraphics{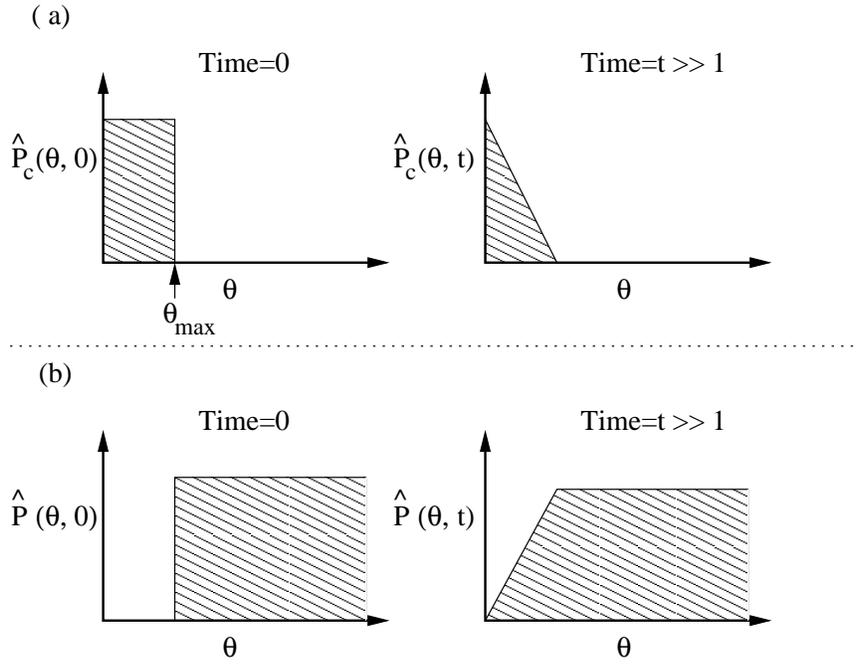}
\caption{The evolution of $\hat{P}_{c}(\theta, t)$ and $\hat{P}(\theta,t)$.}
\label{fig:complement}
\end{figure}

In contrast to the case of the scalloped channel with $180^{o}$ arcs, in the 
other cases shown in Fig.~\ref{fig:examples}, the scattering of a long flight
upon reflection from a channel wall leads to a much more drastic change in the 
angle of a particle's velocity vector with respect to the channel axis.  
For example, for the case of the Sinai billiard, the angular deflection is 
typically of order $\hat{\theta}^{1/2}$ which , for small $\hat{\theta}$, is 
much larger than $\hat{\theta}$.  For the case of the scalloped channel with 
arcs of less than $180^{o}$, a particle moving nearly parallel to a channel axis
is scattered by an angle of order one.  Furthermore, after a large deflection, 
the orientation of the particle's velocity vector is rapidly randomized by a 
succession of many reflections which, since the particle is no longer in a 
long flight, can occur in a relatively short time.  These considerations lead
us to a model for the cases in 
Figs.~\ref{fig:examples}(a)-\ref{fig:examples}(c) and \ref{fig:examples}(d2)
in which we adopt the model hypothesis that, when a particle in a long flight 
suffers a collision with a billiard wall, the orientation of its velocity 
vector is randomly scattered with uniform probability density in $[0,2\pi]$.
We wish to determine the evolution from the initial condition (a) in 
Fig.~\ref{fig:initDist} for the case $\theta_{\textrm{max}} \ll 1$.
This initial distribution is equal to the initial distribution (b) in 
Fig.~\ref{fig:initDist} minus the initial distribution (c) in 
Fig.~\ref{fig:initDist}.  The initial condition (b), which is uniformly 
distributed in angle, remains unchanged when it is evolved forward in time 
$[\hat{P_{b}}(\theta,t)=\hat{P_{b}}(\theta,0)]$, since it is an invariant 
distribution.
Thus, to find the evolution from initial condition (a), we can determine the 
evolution from (c), and then subtract it from (b).
The long time evolution from (c) can be found by considering the time at 
which particles are scattered.  Consider, for example, the scalloped channel,
Fig.~\ref{fig:examples}(d2), and a particle with a small initial $\theta_{o}$.
Suppose the particle is located in the channel at a distance $\Delta x$
from the boundary of the channel with which it will collide [left or right
vertical dashed line in Fig.~\ref{fig:examples}(d2)].  If 
$\Delta x < vt \sin \theta_{o} \cong vt\theta_{o}$, the particle
scatters; if $\Delta x > vt\theta_{o}$, it does not scatter.  Since the 
particles we are considering are in the channel, $\Delta x <W$, every
particle with $\theta_{o}>W/(vt)$ must have scattered at least once.
We assume that $t >W/(v\theta_{max}) \equiv t_{o}$.  Since 
$\theta_{\textrm{max}}$ is small, the scattered particles contribute a
small positive value of order $\theta_{\textrm{max}}$ to 
$\hat{P}_{c}(\theta,t)$ in $0 \leq \theta \leq 2\pi$.  Thus 
$\hat{P}_{c}(\theta,t)$ is small (i.e., of order $\theta_{\textrm{max}}$)
for $\theta > W/vt$.  For $\theta_{o}< \Delta x/(vt)$, $t>t_{o}$, the 
particle has not yet scattered.  Assuming that the initial spatial 
distribution of particles in the channel is uniform, the fraction of 
particles with initial angle $\theta_{o}$ that have scattered is 
$\theta_{o}vt/W$.  Thus

\begin{equation}
\hat{P}_{c}(\theta,t) \cong \left\{
\begin{array}{ll}
\hat{P}_{c}(\theta, 0)(1-\theta vt/W) & \textrm{ for } \theta< W/(vt),\\
0 & \textrm{ for } \theta > W/(vt)
\end{array} \right. ,
\end{equation}
where we have neglected the small, order $\theta_{max}$, contribution to 
$\hat{P}_{c}(\theta,t)$ from scattered particles.  Subtracting $\hat{P}_{c}$
from $\hat{P}_{b}$ as illustrated in Fig.~\ref{fig:complement}, we obtain the 
time asymptotic form in Fig.~\ref{fig:sinai} and Eq.~(\ref{eq:Ptilde}).

\section{Discussion}

As mentioned in the introduction, 
Bleher [9] proved, up to some natural conjectures that ``For any periodic 
configuration of scatterers with an infinite horizon the limit in distribution
\begin{equation}\label{eq:bleher}
\lim_{t\to \infty}\frac{r(t)-r(0)}{(t\log t)^{1/2}}=\eta
\end{equation}
exists and $\eta$ is a Gaussian random variable.'' (We have substituted 
our $r(t)$ for Bleher's $\mathbf{x}(t)$ for the sake of notational 
consistency).

We take no issue with Bleher's  proof referred to above, but we emphasize its 
use of \textit{limit in distribution} convergence.
This kind of convergence is the least restrictive kind of convergence 
considered within the context of probability and statistics.  A proof 
of convergence in distribution implies only the ability to calculate the
expectation value of functions that remain bounded.
The convergence required by a limit in distribution is only strong enough 
to allow calculation of the expectations of functions that remain bounded [10]. 
Therefore the convergence of Bleher's pdf is not strong enough to allow 
the moments of the distribution to be calculated.  As a result of the strong 
weight $|r|^{q}$ puts on the tails of the distribution $(|r| \to \infty)$, 
two different distributions with the same limit in distribution can have 
very different moments.  

The fact that Bleher is able to accurately (according to our simulations) 
calculate the second and lower moments of the displacement distribution 
suggests that his result can be strengthened to ``convergence in $q^{th}$ 
mean''  which is satisfied for a sequence $X_{n}$ if the expectation value of 
$|X_{n}-X|^{q} \to 0$ as $n\to \infty$.  Convergence in $q^{th}$ 
mean also implies that the expectation value of $|X_{n}|^{p}$ limits to 
the expectation value of $|X|^{p}$ for $1 \leq p \leq q$ [15].  Thus 
our results are consistent with convergence in $q$th mean to Bleher's 
distribution for $q=2$, but rule out convergence for any higher value of $q$.

The inapplicability of Bleher's result explains the discrepancy between 
Eqs.~(\ref{eq:moment_intro})-(\ref{eq:gamma}) and the result 
$\langle |r|^{q} \rangle \sim (t \log t)^{q/2}$ one would find 
by mistakenly calculating moments using Bleher's pdf.
Equations~(\ref{eq:moment_intro})-(\ref{eq:gamma}) also differ from normal 
diffusion, $\langle |r|^{q} \rangle \sim t^{q/2}$, as well as from the 
result suggested in [11]. 

In conclusion, our two main results are as follows:
\begin{description}
\item[ ](a) The moments $\langle r^{q}\rangle$ scale as $t^{\gamma_{q}}$ with 
$\gamma_{q}$ given by Eq.~(\ref{eq:gamma}) for any initial bounded 
distribution,$|P(x,y,\theta,0)|<K$, that is zero outside some finite region 
(in particular, Eq.~(\ref{eq:gamma}) still applies if the initial distribution
has no particles with infinite flights).
\item[ ](b) If the initial distribution has no particles in a 
$\theta-$interval 
about a direction of infinitely long flight (say $\theta=0$), then 
$\hat{P}(\theta, t)=\int_{\mathcal{R}}P(x,y,\theta,t)$ approaches a 
time-invariant scaling form $\tilde{P}(\phi)$, where $\phi=\theta t$ and 
$\tilde{P}(\phi)$ is universally the same (Fig.~\ref{fig:sinai}) for the 
billiards of Figs.~\ref{fig:examples}(a),\ref{fig:examples}(b),
\ref{fig:examples}(c), \ref{fig:examples}(d2), but is different 
(Fig.~\ref{fig:spread})
for the billiard of Fig.~\ref{fig:examples}(d1).
\end{description}

This work was supported by the Office of Naval Research (Physics) 
and by the National Science Foundation (award DMS 0104087).  
We thank L. Bunimovich and J. R. Dorfman for discussion.

\section{References}
\noindent a.  Department of Physics and Institute for Research in 
Electronics and Applied Physics.

\noindent b.  Department of Mathematics and Institute for Physical
Science and Technology.

\noindent c.  Department of Electrical and Computer Engineering.

\noindent 1. H. A. Lorentz, Proc. Amst. Acad. \textbf{7}:438, 585, 604
(1905)

\noindent 2. E. H. Hauge, \textit{Lecture Notes in Physics,} Vol. 31, p. 337. (1974).

\noindent 3. G. Gallavotti, \textit{Lecture Notes in Physics,} Vol. 38, p 236 (1975).

\noindent 4. Ya. G. Sinai, \textit{Funkts. Anal. Ego Prilozh.} \textbf{13}:46 (1979).

\noindent 5. L. A. Bunimovich and Ya. G. Sinai, \textit{Commun. Math. Phys.} \textbf{78}:479 (1981).

\noindent 6. P. Castiglione, A. Mazzino, P. Muratore-Ginanneschi, A. Vulpiani, 
\textit{Physica D} \textbf{134}:75 (1999).

\noindent 7. R. Ferrari, A. J. Manfroi, W. R. Young, \textit{Physica D} \textbf{154}:111 (2001).

\noindent 8.   The numerical simulations of the four examples were carried out
in a single cell, such as the cells shown in Fig.~\ref{fig:stadium}.  The 
path of a particle in the infinite domain billiard is extracted by reflection
of the particle trajectory at each straight wall of the cell.  For infinite 
domain billiards with random displacements, the displacement from the center 
of the cell was altered every time the particle reflected from a straight wall
(i.e. entered a new cell).  For simplicity we 
altered the displacement even when the particle entered a previously visited 
cell; thus the simulation differs slightly from the fixed scatterer scenario, 
but the discrepancy affects only lower order terms.
Likewise the orientation of the square was altered
every time the particle reflected from a straight wall.

\noindent 9. P. M. Bleher \textit{J. Stat. Phys.} \textbf{66}:315 (1992).

\noindent 10. A. Papoulis, \textit{Probability, Random Variables, and
Stochastic Process,} 4th ed. (McGraw-Hill, 1994).

\noindent 11. G.M. Zaslavsky and M. Edelman, Phys. Rev. E 
\textbf{56}:5310 (1997).

\noindent 12. K.C. Lee, Phys. Rev. Lett. \textbf{60}:1991 (1988).

\noindent 13. D. N. Armstead B. R. Hunt and E. Ott, to be published (2002).

\noindent 14. P. Dahlqvist, J. Stat. Phys. \textbf{84}:773 (1996).

\noindent 15. P. Pollett MS308 Probability Theory web page, \\
http://www.maths.uq.edu.au/~pkp/teaching/ms308/summaries/conv/conv.html.

\end{document}